\documentclass[osajnl,twocolumn,showpacs,superscriptaddress,10pt]{revtex4-1}
\usepackage{amsmath,amssymb,graphicx}

\def\Tr{\hbox{Tr}}

\def\bmsigma{\boldsymbol{\sigma}}
\def\bmSigma{\boldsymbol{\Sigma}}

\def\bmA{\boldsymbol{\Sigma}}
\def\bmB{\boldsymbol{\Sigma}}
\def\bmC{\boldsymbol{\Sigma}}
\def\bmzero{\boldsymbol{0}}
\def\bmdelta{\boldsymbol{\delta}}
\def\Minfo{{\mathcal I}}

\def\PRL{{Phys. Rev. Lett.} }

\newcommand{\avefr}[1]{{\langle #1 \rangle_{\rm fr}}}
\begin{document}
%\twocolumn[ %% activate for two-column option
\title{Revealing interference by continuous variable discordant states}
\author{A.~Meda}\email{Corresponding author: a.meda@inrim.it}
\affiliation{INRIM, Strada delle Cacce 91, I-10135
  Torino, Italy}
\author{S.~Olivares}
\affiliation{Dipartimento di
  Fisica, Universit\`a degli Studi di Milano, I-20133 Milano, Italy}
  \author{I.~P.~Degiovanni}
\author{G.~Brida}
\author{M.~Genovese}
\affiliation{INRIM, Strada delle Cacce 91, I-10135
  Torino, Italy}
\author{M.~G.~A.~Paris}
\affiliation{Dipartimento di Fisica, Universit\`a degli Studi di Milano, I-20133 Milano,
Italy}
\begin{abstract} In general, a pair of uncorrelated Gaussian states mixed in a beam
splitter produces a correlated state at the output. However, when the
inputs are identical Gaussian states the output state is equal to the input,
and no correlations appear, as the interference had not taken place.
On the other hand, since physical phenomena {\em do have} observable effects, and
the beam splitter is there, a question arises on how to reveal the
interference between the two beams. We prove theoretically and demonstrate
experimentally that this is possible if at least one of the two
beams is prepared in a discordant, i.e. Gaussian correlated, state
with a third beam. We also apply the same technique to reveal the erasure
of polarization information. Our experiments involves thermal states and
the results show that Gaussian discordant states, even when they show a
positive Glauber P-function, may be useful to achieve specific tasks.
\end{abstract}
\ocis{(270.5290) Photon statistics; (270.0270) Quantum optics;
(270.6570)  Squeezed states}
\maketitle %% required
Understanding the nature of correlations among quantum systems is one of
the major task of current research. Quantum correlations, in fact, play
a leading role in understanding the very foundations of quantum
mechanics, and represent the basic resource for the development of
quantum technologies. Different quantities and strategies to
discriminate whether correlations have a quantum nature or not
%\cite{modi:rev,1,2,3,4,5}
\cite{modi:rev,1,3} have been introduced, and it has also been
pointed out \cite{ferr:PRL:12,vog} that the criteria based on the
informational point of view, such as the quantum discord
%\cite{OZ:disc,HV:disc,GQD10,Ade10,QD11a,QD11b,expD2,expD3,d}
\cite{OZ:disc,HV:disc,GQD10,Ade10,QD11b,expD2,expD3,d}, are
somehow incompatible with the physical ones based on the
Glauber-Sudarshan phase-space approach 
%\cite{Gla63,Sud63,Gla69}.
\cite{Gla63,Sud63}.
A paradigmatic example in quantum optics is given by a thermal
equilibrium state divided at a beam splitter (BS). This state, which
is characterized by Gaussian Wigner functions, is indeed a classical
one according to the Glauber approach, however, the bipartite state
emerging from the BS displays non-zero Gaussian discord and, thus, from the
informational point of view it contains a non-vanishing amount of
quantum correlations. It is also worth noting that, for Gaussian
states, the only bipartite states with zero Gaussian discord are the
factorized ones \cite{GQD10,ralph12} and that there are evidences that
the Gaussian discord could be the ultimate quantum discord for
Gaussian states \cite{Ade10,ng2}.
\begin{figure}[tpb]
\centering
\includegraphics
%[width=8.4 cm, height=7 cm, bb=0 0 400 250]{f1_setup.jpg}
[width=8.4 cm, height=4 cm]{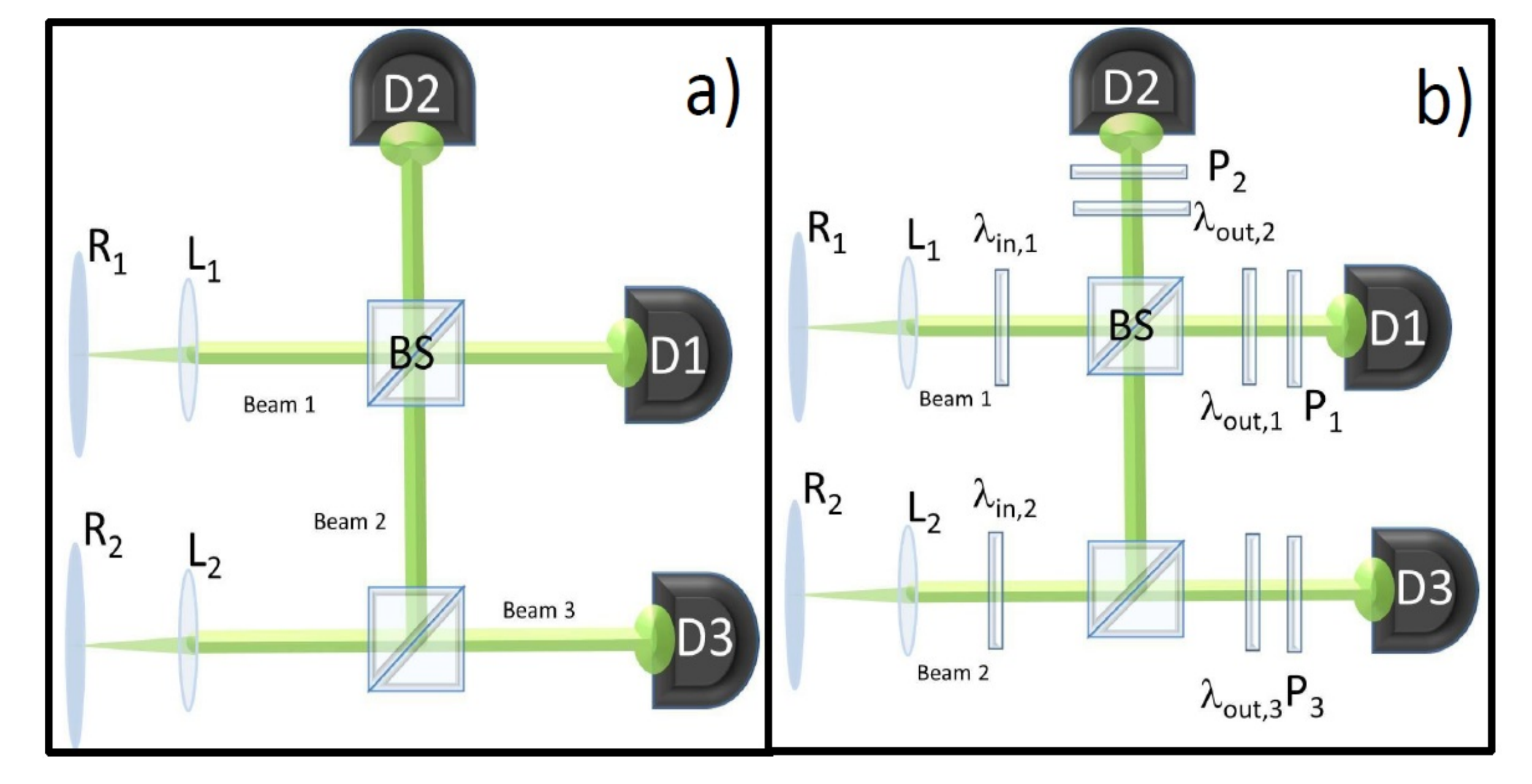}
\caption{Revealing interference by continuous variable discordant
states:scheme of the two experimental setups.}
\label{Setup}
\end{figure}
In general, if a factorized state $\varrho_{12} = \varrho_1 \otimes
\varrho_2$ undergoes a unitary interaction described by the operator
$U_{12}$, then the evolved state $\tilde{\varrho}_{12} =
U_{12}\varrho_{12}U_{12}^{\dag}$ may be correlated. The total amount
of correlations can be quantified by the mutual information:
$\Minfo[\tilde{\varrho}_{12}] = S[\tilde{\varrho}_{1}] +
S[\tilde{\varrho}_{2}] - S[\tilde{\varrho}_{12}] = \Delta S_1 + \Delta
S_2$, where $\tilde{\varrho}_{k} = \Tr_{h}[\tilde{\varrho}_{12}]$,
$h\ne k$, $S[\tilde{\varrho}_{k}]=
-\Tr[\tilde{\varrho}_{k}\ln\tilde{\varrho}_{k}]$ is the von Neumann
entropy and $\Delta S_{k} = S[\tilde{\varrho}_{k}] - S[\varrho_{k}]$.
>From the above equation we can see the rise of correlation as due to
an increase of entropy between the input and output states
$\varrho_{k}$ and $\tilde{\varrho}_{k}$, respectively. It is clear
that if $\varrho_{k} = \tilde{\varrho}_{k}$, then
$\Minfo[\tilde{\varrho}_{12}]=0$ (provided that the input is a
factorized and thus uncorrelated state). For Gaussian states, this
happens when the inputs have the same covariance matrix (CM) and
$U_{12}$ corresponds to a bilinear, energy-conserving interaction
described by $H_I \propto a^\dag b + a b^\dag$, where $a$ and $b$ are
bosonic annihilation operators, $[a,a^\dag]=1$ and $[b,b^\dag]=1$
%\cite{kim:09,oli:09,inv11,oli:11}
. 

%These Hamiltonians describe the
%interaction of two light modes in a BS or a frequency converter, but
%also other quantum systems of interest for quantum technology, e.g.,
%collective modes in a gases of cold atoms \cite{meystre}, atom-light
%nondemolition measurements \cite{Nat11}, optomechanical oscillators
%\cite{pir:03,xia:10}, nanomechanical oscillators \cite{cav08}, and
%superconducting resonators \cite{chi:10}.  

When the initial state
$\varrho_{12}$ and the evolved one $\tilde{\varrho}_{12}$ are exited
in the same factorized state, they cannot be discriminated and no
correlations appear, as the interference of the two beams had not
taken place.  On the other hand, since physical phenomena {\em do
  have} observable effects, and the BS is there, a question arises on
how to reveal the interference between the two beams.
In this letter we investigate the dynamics of correlations in this
kind of systems and demonstrate, both theoretically and
experimentally, that revealing interference is possible by adding an
ancillary mode 3 correlated with one of the two beams, say beam
2. More explicitly, it is sufficient that the bipartite state
$\varrho_{23}$ has non zero Gaussian discord to reveal the
interference between mode 1 and 2 even when the local states
$\varrho_2 = \hbox{Tr}_3[\varrho_{23}]\equiv\varrho_1$ are identical
and the interaction at the BS is not creating any correlations between
them.
Let us consider two generic zero-amplitude Gaussian states \cite{nota1} $\varrho_k = S(r_k)
\nu_{\rm th}(N_k)S^{\dag}(r_k)$, where $\nu_{\rm th}(N_k) =
\sum_{n=0}^{\infty}(N_k)^{n}/(1+N_k)^{n + 1} | n \rangle \langle n |$
is a thermal equilibrium state with $N_k$ thermal photons and
$S(r_k) = \exp\{\frac12 r_k [ (a_k^{\dag})^2 - a_k^2]\}$
is the squeezing operator, $a_k$ being mode operators, $k=1,2$.
The $2\times 2$ CM of the state $\varrho_k$
can be written as $\bmsigma_{k} \equiv
\bmsigma(N_{{\rm tot},k},\beta_k)$, where
$\bmsigma(N,\beta) = \hbox{Diag}\left\{
f_{+}(N,\beta), f_{-}(N,\beta)\right\}$,
$f_{\pm}(N,\beta) = \frac12+N \pm \sqrt{\beta N[1+N(2-\beta)]}$
and we introduced the total number of photons $N_{{\rm tot},k} =
\Tr[a_k^{\dag} a_k\,\varrho_k]$ and the squeezing fraction $\beta$,
whith $N_k = (1-\beta) N_{{\rm tot},k} $. We have
assumed $r_k>0$ without loss of generality.
With this notation, $\beta =
0$ and $\beta=1$ correspond to the thermal and the squeezed vacuum
state, respectively, while $\bmsigma(0,0) \equiv \bmsigma_0$ is the CM
of the vacuum state $\varrho_{0} = | 0 \rangle \langle 0 |$.
Under the action of a BS with transmissivity $\tau$, the initial
$4\times 4$ CM $\bmSigma_0 = \bmsigma_1\oplus \bmsigma_2$ of the
two-mode state $\varrho_1\otimes \varrho_2$ transforms as
$
\bmSigma_0 \to \bmSigma^{\rm (out)} = \left(\begin{array}{cc} \bmA_1
& \bmC_{12}\\ [1ex] \bmC_{12} & \bmB_2
\end{array}\right),
$
where $\bmA_1 = \tau\bmsigma_1 + (1-\tau)\bmsigma_2$, $\bmB_2 =
\tau\bmsigma_2 + (1-\tau)\bmsigma_1$ and $\bmC_{12} =
\tau(1-\tau)(\bmsigma_2-\bmsigma_1)$. Note that $\bmC_{12} \ne
\bmzero$ denotes the presence of correlation between the outgoing
modes. Notice that rewriting $\bmC_{12} = \tau(1-\tau)[(\bmsigma_2 -
\bmsigma_0) + (\bmsigma_0 -\bmsigma_1)]$, we can identify two
different contributions: the one, $\propto (\bmsigma_0 -\bmsigma_1)$,
which is equal to that obtained by mixing $\varrho_1$ with the vacuum,
i.e., $\varrho_{2} \equiv \varrho_{0}$; similarly, the other, $\propto
(\bmsigma_2 -\bmsigma_0)$, corresponds to that obtained by mixing
$\varrho_2$ with $\varrho_{1} \equiv \varrho_{0} $. On the other hand,
interference cannot be seen as the simple sum of two contributions and
this will be exploited later on in this letter in order to describe
the results of our second experiment.
As follows from the above analysis, if the input modes are prepared in
the same initial state, i.e., $\bmsigma_1 = \bmsigma_2$, then the
output beams are left in an uncorrelated, factorized state with
$\bmSigma_0 \equiv \bmSigma^{\rm (out)}$ and ($\bmC_{12} =
\bmzero$). In this case the two above-mentioned contributions cancel
each others and the interaction leaves the system unchanged.  In order
to reveal inteference, we correlate mode 2 with a third auxiliary mode
3, i.e., we prepare $\varrho_{23} \ne \varrho_2 \otimes \varrho_3$
such that $\varrho_2 = \hbox{Tr}_3[\varrho_{23}] = \varrho_1 =
\varrho$. Modes 1 and 2 are still left unchanged and uncorrelated
after the interference, but now, because of the interaction, part of
the correlations shared between modes 2 and 3 are now shared between
modes 1 and 3. This monogamy effect \cite{Gmo} can be seen by looking
at the evolved CM of the whole state of the three modes.  The $6\times
6$ CM of the initial state $\varrho_{123} = \varrho_1\otimes
\varrho_{23}$ reads: $ \bmSigma_{123} = \bmsigma_1 \oplus \left(
\begin{array}{cc}
\bmsigma_2 & \bmdelta_{23} \\
\bmdelta_{23}^T & \bmsigma_{3}
\end{array}
\right)$ where $\bmsigma_k$ is the $2\times 2$ single-mode CM of mode
$k=1,2,3$, $\bmsigma_1 = \bmsigma_2 = \bmsigma(N,\beta)$, $N$ being
the total number of photons per mode. The block $\bmdelta_{23} \ne
\bmzero$ contains the correlations between modes 2 and 3, which show
nonzero Gaussian A- and B-discord \cite{ijmpb12}.
%%%%%%%%%
After mixing mode 1 and 2 at the BS we have:
\begin{equation}
\bmSigma_{123} \to \bmSigma_{123}^{\rm (out)} =
\left(\begin{array}{ccc}
\bmsigma(N,\beta) & \bmzero & \sqrt{1-\tau}\, \bmdelta_{23}\\
\bmzero & \bmsigma(N,\beta) & \sqrt{\tau}\, \bmdelta_{23}\\
\sqrt{1-\tau}\, \bmdelta_{23} & \sqrt{\tau}\, \bmdelta_{23} &
\bmsigma_3
\end{array}
\right).\label{CM:out}
\end{equation}
The comparison between input and output CMs shows that while modes 1
and 2 are (locally) left unchanged and uncorrelated, both of them are
now correlated with mode 3. Furthermore, the degree of correlations
between the modes 2 and 3 is decreased ($\bmdelta_{23} \to
\sqrt{\tau}\, \bmdelta_{23}$) for the benefit of the birth of
correlations between the previously uncorrelated modes 1 and 3
($\bmzero \to \sqrt{1-\tau}\, \bmdelta_{23}$).
It is worth noting that
the birth (reduction) of correlation between modes 1 and 3 (modes 2
and 3) is not merely due to the transmission (reflection) of beam 2,
but it is due to its interference at the BS: beam 2 evolves in a
two-mode correlated state, whose modes are thus correlated with mode
3. For the sake of clarity, we addressed only single-mode beams, but
the same results hold also in the presence of multimode Gaussian beams
since the phenomenon is essentially due to the tensor product nature
of the multimode state, and to the pairwise nature of the interaction
at the BS.
In the experiment, we exploit correlations among three spatial multimode pseudo-thermal beams. We
produce two independent unpolarized beams with thermal statistics
addressing $1$~ns laser pulses at $532$ nm on two independent rotating
ground glasses R1 and R2, with inhomogeneities of approximately
1~$\mu$m of size.  The two speckled beams are collimated with two
lenses ($L_1$ and $L_2$) of $f = 1.5$ m focal length put at a
distance $f$ from the disks. Beam 1 is directly sent to a balanced BS
while the second is further divided into beams 2 and 3
(Fig.~\ref{Setup}a). Each beam $k=1,2,3$, is then sent to the
corresponding detector D$k$, which is a portion of a CCD sensor. The
speckled beams are imaged by means of a lens of focal lens $f_I = 25$
cm on the array of pixels. Due to the presence of the lenses $L_1$ and
$L_2$, each speckle on the CCD array corresponds to a spatial mode of
the pseudo-thermal beam. For each beam $k$ we select an area $A_k$
collecting $M$ spatial modes and evaluate the intensity
$I_k^{(j)}=\sum\limits_{m=1}^M \langle a_{m,k}^\dag
a_{m,k} \rangle$ for each frame $j$ of the CCD where
$a_{m,k}$ is the field operator of the $m$-th mode impinging on
the area $k$. The correlation between the beams $h$ and $k$ is
estimated by using the second order correlation coefficient
$
c_{h,k}=\frac{ \avefr{I_k I_h} -
\avefr{I_h}\avefr{I_k}}{\Delta_{\rm fr}(I_h)\Delta_{\rm fr}(I_k)}
$,
where
$\avefr{F}=(N_{\mathrm{frame}})^{-1}\sum_{j=1}^{N_{\mathrm{frame}}}F^{(j)}$
is the average over $N_{\mathrm{frame}}$ frames and $\Delta_{\rm
  fr}(I_k)^2=\avefr{I_k^2}-\avefr{I_k}^2$.  It is worth noting that
$c_{h,k}$ is independent on the number of modes $M$, provided that all
spatial modes of each beam have the same intensity.
In order to align the setup, and to achieve the proper mode matching
at the BS, we first realize the superposition of the correlated areas
$A_1$ and $A_2$ by alternatively stopping beam 1 or beam 2 and maximizing
the correlation $c_{1,2}^{(1)}$ and $c_{1,2}^{(2)}$ between the beams
outgoing the BS. We obtain $c_{1,2}^{(1)} = 0.97$ and
$c_{1,2}^{(2)}=0.96$.

\begin{table}[tbh]
\centering
\setlength{\tabcolsep}{10pt}
\begin{tabular}{c c c}
beams $h,k$ & $c_{h,k}^{(in)}$ & $c_{h,k}^{(out)}$  \\[1ex]
  \hline\hline
$1,2$ & $0.09\,_{[-0.28;0.46]}$  & $-0.01\,_{[-0.38;0.35]}$ \\[1ex]
\hline
$1,3$ & $-0.01\,_{[-0.38;0.36]}$ & $0.55\,_{[0.29;0.81]}$ \\[1ex]
\hline
 $2,3$ & $0.97\,_{[0.85;1.00]}$ & $0.62\,_{[-0.38;0.85]}$ \\[1ex]
 \hline\hline
\end{tabular}
\caption{Measured correlations between beams $h,k$ before ($in$) and
  after ($out$) the BS. The subscripts report the
confidence intervals at $99\%$.} \label{tab1}
\end{table}

\begin{figure}[tpb]
\centering
\includegraphics [width=8.4 cm, height=3.5 cm]{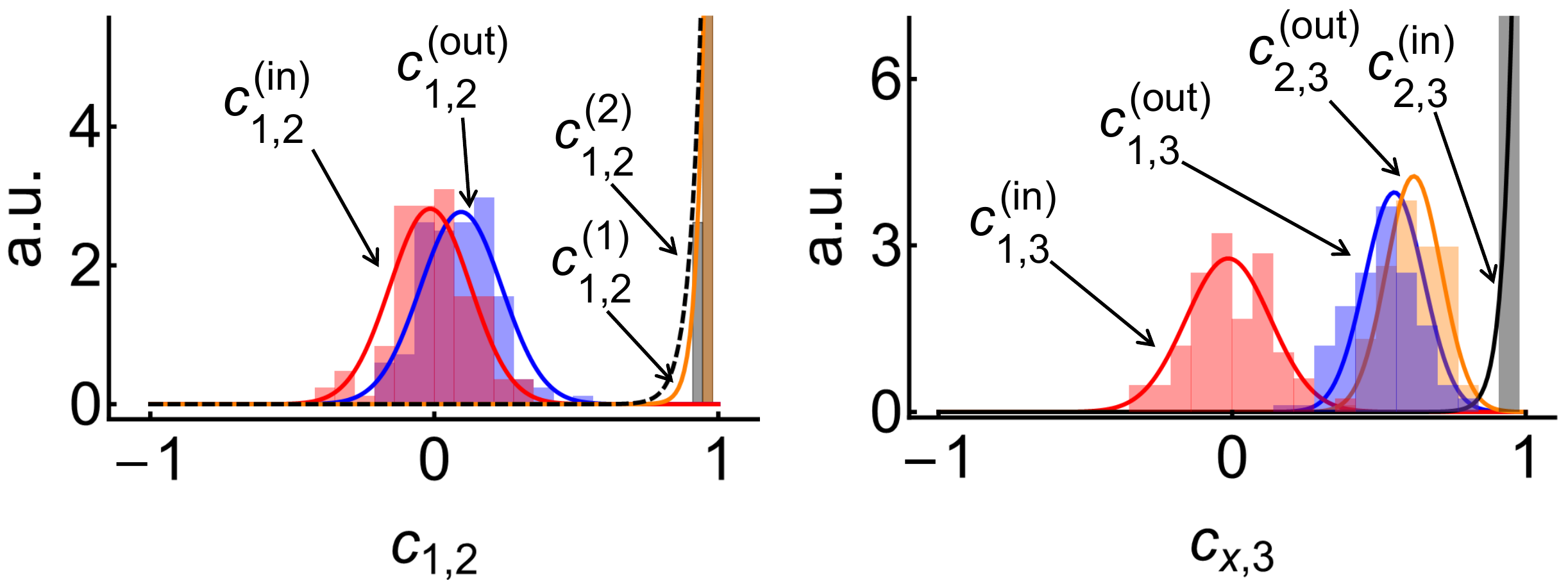}
 \caption{Correlation coefficients among the three beams before and
 after mixing of modes 1 and 2. The first plot shows the evolution of
 the correlation between beams 1 and 2. The second
 plot refers to correlations among beam 1 and 2 with beam 3.}\label{f:corr}
 \end{figure}

We then measure the correlations coefficients $c_{1,2}^{(in)}$,
$c_{1,3}^{(in)}$ and  $c_{2,3}^{(in)}$ of the initial
state and $c_{1,2}^{(out)}$, $c_{1,3}^{(out)}$ and $c_{2,3}^{(out)}$
of the states after the mixing of beam 1 and beam 2 in the BS.
Experimental results are summarized in Fig. \ref{f:corr} and
in Tabel~\ref{tab1}, where we report the measured correlations between the
couples of beams before and after the interaction with the BS
averaging over $N_{\rm frame}=50$ frames. The mean values and the
confidence intervals (at $99\%$) are obtained from the raw data by taking into
account the bounded nature of the correlation coefficients
$c_{h,k}$ \cite{lir:09}.
As it is apparent from Fig. \ref{f:corr} and from Table \ref{tab1}, beams 1 and 2
are not affected by the presence of the BS (the small discrepancies between
the measured correlation are due to the slightly imperfect mode matching),
whereas the interference between them is revealed by the dramatic change
in the correlations with mode 3.

%(pezzo aggiunto)

\begin{figure}[tbh]
\centering\label{f:f3_cvsd}
\includegraphics[width=0.48\columnwidth,angle=0]{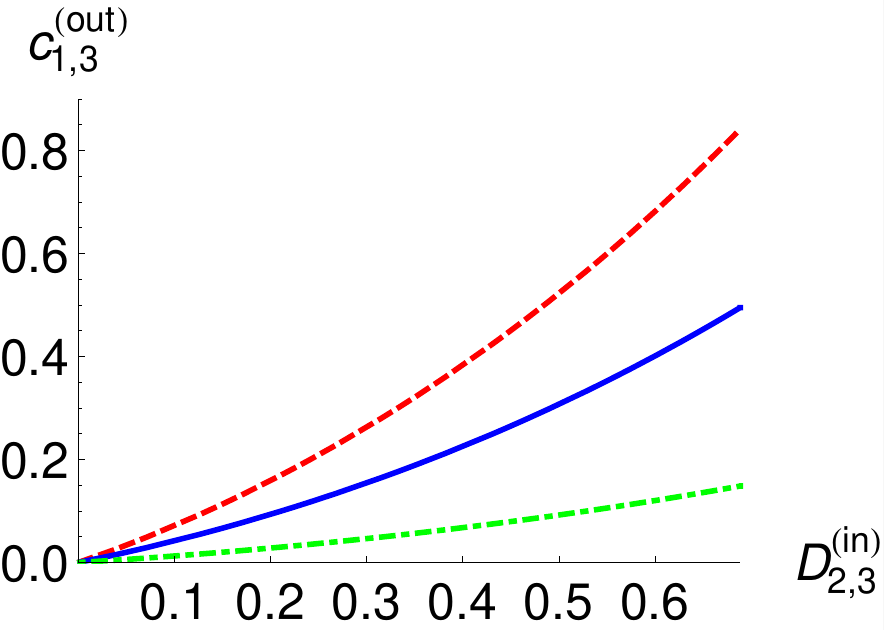}
\includegraphics[width=0.48\columnwidth,angle=0]{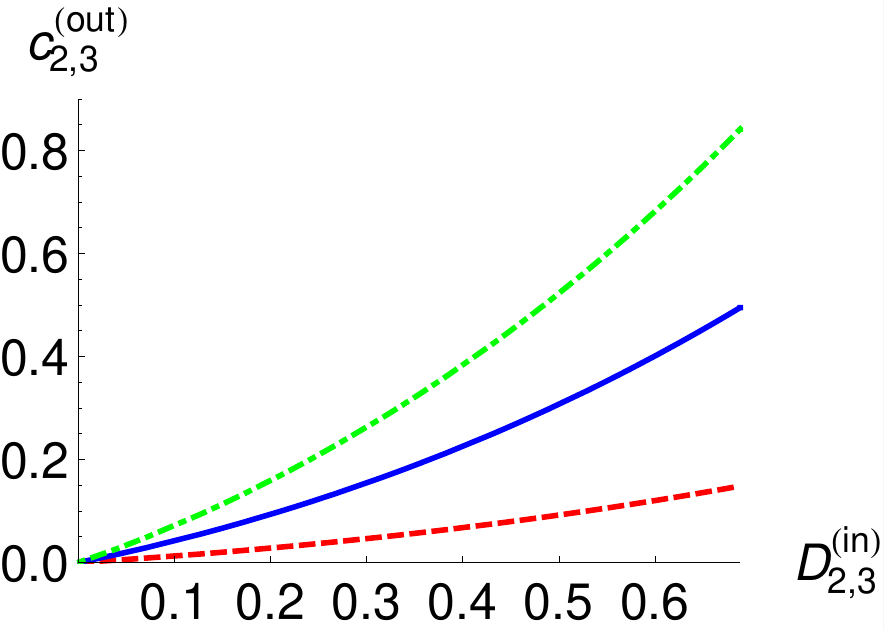}
\caption{(Color online) Output correlations $c_{1,3}^{(out)}$ (left) and
$c_{2,3}^{(out)}$ (right) as a function of the initial discord
between modes 2 and 3. In both panels the red dashed lines denote the
curves for transmissivity equal to 15\%, the solid blue lines are for
the balanced case and the green dot-sashed lines for transmissivity
85\%.}\label{f:f3}
\end{figure}

The significant role of discord in our
protocol is illustrated in Fig. \ref{f:f3}, where we show the behavior
of the output correlations $c_{1,3}^{(out)}$ and $c_{2,3}^{(out)}$ as a function
of the discord between modes 2 and 3 at the input.
As it is apparent from the plots, correlations at the output are
monotone functions of the initial discord. Nonzero
correlations are created for any value of initial discord.
The three lines in both panels of Fig. \ref{f:f3} correspond
to three different values of the transmissivity of the beam splitter
creating discord between modes 2 and 3. As expected from the
form of the CM in Eq. (\ref{CM:out}), increasing the transmissivity
of the beam splitter increases the output correlations between
modes 2 and 3 at the expense of correlations between modes 1 and 3.
%%%%%%%%%%%%%%%%%%%%%%%%%%%%%%%%%%%%%%%%%
\par
%%%%%%%%%%%%%%%%%%%%%%%%%%%%%%%%%%%%%%%%%
%(FINE pezzo aggiunto)

In order to further clarify the role of the ancillary mode 3 we now
consider a different scenario, where the two input beams do not
interact at the BS. As depicted in Fig.~\ref{Setup}b, this is
achieved by two half wave plates $\lambda_{in,1}$ and $\lambda_{in,2}$,
which set horizontal polarization ($H$) for beam 1, i.e.,
$\varrho^{(H)}_{1}$, and vertical polarization ($V$) for beam 2,
$\varrho^{(V)}_{2}$.  We assume that mode 2 and 3 have the same
polarization. Due to the different polarizations, modes 1 and 2 no longer
interfere at the BS: rather, they both interact with a vacuum mode
with the same polarization entering the other port of the beam, thus
giving rise to two couples of collinear, superimposed correlated beams
one with $V$ polarization, the other with $H$ polarization. Overall,
we have four modes, and the two states at the output are
distinguishable.  If we put two polarization filters after the BS, we
can select beams with a fixed polarization $\alpha = H,V$, which are
Gaussian states with CM given by (we set $\tau=1/2$):
%\begin{align}
$\bmSigma_{\rm out}^{(H)} = \frac12 \left(\begin{array}{cc}
\bmsigma_1^{(H)} +\bmsigma_{0} & \bmsigma_{0}-\bmsigma_1^{(H)}\\ [1ex]
\bmsigma_{0}-\bmsigma_1^{(H)} & \bmsigma_1^{(H)} +\bmsigma_{0}
\end{array}\right)
$ and 
%,\\[2ex]
$\bmSigma_{\rm out}^{(V)} = \frac12 \left(\begin{array}{cc}
\bmsigma_2^{(V)} +\bmsigma_{0} & \bmsigma_2^{(V)}-\bmsigma_{0}\\ [1ex]
\bmsigma_2^{(H)} -\bmsigma_{0}& \bmsigma_2^{(V)} +\bmsigma_{0}
\end{array}\right),
$
%\end{align}
respectively, where $\bmsigma^{(\alpha)}_{k}$ are the same as
$\bmsigma_{k}$, $k=1,2$, but now we emphasize the polarization
dependence $\alpha=H,V$. Thanks to the polarization, we
can clearly distinguish the correlations coming from the off diagonal
$\propto (\bmsigma_0 - \bmsigma_1^{(H)})$ and $\propto
(\bmsigma_2^{(V)} -\bmsigma_0)$.
In this experiment, the physical action that we want to reveal
is the {\em erasure} of the information about the
polarization. This is done as in the quantum erasure protocol for
discrete variables \cite{kwi:92}: we insert two polarization rotators
set at 45$^{\circ}$ after the BS and on the path of mode 3.  After
filtering, the resulting three $H$-polarized ($V$-polarized) modes
have the same CM as in Eq.~(\ref{CM:out}) for a suitable choice of the
input total energy and squeezing fraction.
We measured the second order correlation coefficient
between the two beams before and after the BS without acting on their
polarizations, obtaining $c_{1,2}^{(H,V,in)} = -0.01$
and $c_{1,2}^{(H,V,out)}= 0.97$, respectively. In this
case, because of the orthogonal polarizations, the beams do not
interfere each other, and each input is divided into two correlated
parties. After the interaction, all the beams are projected to the
45$^{\circ}$ polarization basis by means of three half wave plates
$\lambda_{out,k}$ and three polarizers $P_k$ oriented in the $H$
direction, $k=1,2,3$ (see Fig.~\ref{Setup}b). Again, we measure
correlation $c_{1,2}^{(out)}(\mbox{@} 45^\circ)$,
$c_{1,3}^{(out)}(\mbox{@} 45^\circ)$, and $c_{2,3}^{(out)}(\mbox{@}
45^\circ)$ between the corresponding beams. We then perform the same
measurement projecting the modes onto the vertical basis removing the
half wave-plates [$c_{1,2}^{(out)}(\mbox{@} V)$,
$c_{1,3}^{(out)}(\mbox{@} V)$ and $c_{2,3}^{(out)}(\mbox{@} V)$].
\begin{table}[h!]
\centering
\setlength{\tabcolsep}{10pt}
\begin{tabular}{c c c}
 $c_{1,2}^{(out)}(\mbox{@}45^\circ)$ &
 $c_{1,3}^{(out)}(\mbox{@}45^\circ)$ & $c_{2,3}^{(out)}(\mbox{@}45^\circ)$  \\[1ex]
  \hline\hline
$0.10\,_{[-0.25;0.46]}$  & $0.54\,_{[0.27;0.80]}$  & $0.53\,_{[0.24;0.81]}$ \\[1ex]
 \hline
 \hline
 & & \\
 $c_{1,2}^{(out)}(\mbox{@}V)$ &
 $c_{1,3}^{(out)}(\mbox{@}V)$ & $c_{2,3}^{(out)}(\mbox{@}V)$  \\[1ex]
  \hline\hline
$0.97\,_{[0.87;1.00]}$  & $-0.01\,_{[-0.38;0.36]}$  & $0.97\,_{[0.84;1.00]}$ \\[1ex]
 \hline
 \hline
\end{tabular}
\caption{Measured correlations between the beams $h,k$ after the BS with
  polarizers @~45$^\circ$ and @~$V$. Without polarization selection we have
  $c_{1,2}^{(H,V,in)} =
  -0.01\,_{[-0.38;0.35]}$ and $c_{1,2}^{(H,V,out)}=
  0.97\,_{[0.86;1.00]}$, see text for details.} \label{tab2}
\end{table}
In fact, the erasure of information about polarization affects
correlations between beam 1 and 2 (see Table \ref{tab2}): The correlations
$c_{1,2}^{(H,V,out)}= 0.97$ reduce to $c_{1,2}^{(out)}(\mbox{@}45^\circ) = 0.10$
when the information about initial polarization is lost. Analogously,
beams 2 and 3, which show high correlations in $V$ basis, $c_{2,3}^{(out)}
(\mbox{@}V) = 0.97$, loose correlation in the
45$^{\circ}$ basis [$c_{2,3}^{(out)}(\mbox{@}45^\circ) =
0.53$, while the uncorrelated beam 1 and 3 gains
correlation.  Also in this
case, the use of discordant states for beams 2 and 3 allows to reveal
the physical action, here the erasure, performed on beams 1 and 2,
despite the fact that this cannot be done by inspecting the involved
beams only.
\par
In summary, while a pair of uncorrelated Gaussian
states mixed in a beam splitter produce, in general, a correlated
bipartite state, two equal Gaussian states do not. No correlations
appear at the output, and the interference cannot be detected looking
at the two beams only.  We have proved theoretically and
experimentally that this task may be pursued using an ancillary beam,
prepared in a discordant state with one of the two inferring beams,
thus confirming that discord can be consumed to encode information that can only be
accessed by coherent quantum interactions \cite{Gu12}.
Our experiment involves thermal states and the results show that
Gaussian discordant states, even when they show a positive Glauber
P-function, may be useful to achieve specific tasks.

\par
{\em Acknowledgments}---This research has received
funding from the EU FP7 under grant agreement 308803 (BRISQ2), MIQC,
Fondazione SanPaolo and MIUR (FIRB ``LiCHIS'' - RBFR10YQ3H, Progetto
Premiale ``Oltre i limiti classici di misura'').


\begin{thebibliography}{30}
\bibitem{nota1} Since we are interested in correlations,
we may safely assume that both the input states have zero first moments
without any loss of generality.
\end{thebibliography}
\end{document}